# ProS²Vi: a Python Tool for Visualizing Proteins Secondary Structure

M. Luckman Qasim[1,2], Laleh Alisaraie[1,*]

[1] School of Pharmacy, Memorial University of Newfoundland, 300 Prince Philip Dr, A1B 3V6, St. John's, Canada

[2] Department of Computer Science, Memorial University of Newfoundland, A1C 5S7, St. John's, Canada

*Corresponding author*
*Email: laleh.alisaraie@mun.ca*

## Abstract

The **Pro**tein **S**econdary **S**tructure **V**isualizer (ProS²Vi) is a novel Python-based visualization tool designed to enhance the analysis and accessibility of protein secondary structures calculated and identified by the Dictionary of Secondary Structure of Proteins (DSSP) algorithm. Leveraging robust Python libraries such as "Biopython" for data handling, "Flask" for Graphical User Interface (GUI), "Jinja2", and "wkhtmltopdf" for visualization, ProS²Vi offers a modern and intuitive representation for visualization of the DSSP assigned secondary structures to each residue of any protein's amino acid sequence. Significant features of ProS²Vi include customizable icon colors, the number of residues per line, and the ability to export visualizations as scalable PDFs, enhancing both visual appeal and functional versatility through a user-friendly GUI. We have designed ProS²Vi specifically for secure and local operation, which significantly increases security when dealing with novel protein data.

## 1. Introduction

Assigning secondary structures (SS) to proteins' primary structure (i.e., amino acid sequence) is a fundamental requirement for understanding their biological functions, stability, and intra-molecular as well as inter-molecular interactions. The secondary structures, comprising proteins' amino acids sequence, fold as α-helices, β-strands, β-turns, $3_{10}$ helices, etc., describe the secondary structure and, consequently, protein folding and functionality (Brandt, 2018). Visualization of the two-dimensional (2D) structures of amino acid sequences aids in the interpretation of complex molecular data and in explaining proteins' cellular, biophysical, biomechanical, and biochemical functions so that they can be interpretable to researchers and academic learners alike. Therefore, practical visualization tools are crucial for advancing our understanding of proteins' function as related to their physicochemical properties stemming from their small amino acid building blocks (O'Donoghue, 2021). Visualization tools not only facilitate our understanding of protein structure but also enhance our ability to predict their interactions with other proteins, endo, and exogenous substrates, including therapeutical agents, which are pivotal, for instance, in drug discovery, medicine, and biotechnology (Koch, 2011).

Several algorithms have been developed to assist biochemists and biologists in studying protein secondary structures, each striving for greater accuracy — among the renowned algorithms such as STRIDE (Frishman and Argos, 1995), KAKSI (Martin, et al., 2005), and DSSP (Kabsch and Sander, 1983) — the latter often stands out as the foundation method. Kabsch and Sander introduced the Dictionary of Secondary Structure of Proteins (DSSP) in the early 1980s (Kabsch and Sander, 1983). DSSP has remained a benchmark due to its rigorous methodology based on hydrogen bonding interaction patterns. DSSP classifies protein structures into eight types of secondary structures (e.g., helix, strand, H-bond turn, etc.), offering a detailed view that is critical for understanding protein folding and stability. Comparative studies have been conducted between DSSP and other algorithms like STRIDE and KAKSI, especially concerning irregular peptides and complex structures. DSSP consistently shows robust performance in accurately assigning secondary structures (Zhang and Sagui, 2015). However, the output from these algorithms is often textual and hard to interpret, which can be a significant barrier for those not specialized in bioinformatics. Visualization tools, therefore, play an important role in translating these textual outputs into graphical forms that are easier to understand and interpret.

Web-based visualization tools like STRIDE web-server (Heinig and Frishman, 2004) and POLYVIEW-2D (Porollo, et al., 2004) are valuable for their accessibility and functionality. However, they face challenges that may render them less appealing in the contemporary context of protein research and cybersecurity. One of the main challenges is the security risks associated with these tools. A user is commonly required to upload a protein's structural information to an online web server to generate a 2D diagram to demonstrate the assigned secondary structure of a protein sequence or exhibit its folding in a linear format. Uploading unpublished data of a novel protein structure, newly characterized either experimentally or computationally, to online web servers exposes sensitive data to potential cybersecurity threats, including unauthorized data access, data manipulation, and service disruptions caused by cyber-attacks such as Denial of Service (DoS) (Gu and Liu, 2012). These issues are compounded by the platforms' sporadic security updates, posing a significant risk when handling confidential or proprietary data (Nawaz, et al., 2023). Given these vulnerabilities, researchers must carefully evaluate the security policies and reputational trustworthiness of these web-based tools. For those concerned with data security, alternatives such as local software solutions provide more control over data security (Amo-Filva, et al., 2021). To address the security concerns associated with web-based visualization tools for protein research and study its folding and three-dimensional (3D) structure, we have developed a new local visualization tool using Python as a protein secondary structure visualizer (ProS$^2$Vi) (Qasim and Alisaraie, 2024).

ProS$^2$Vi (Qasim and Alisaraie, 2024) is a tool designed to operate entirely on the user's local machine, which significantly mitigates the risks of data breaches, data integrity attacks, and service disruptions commonly associated with online platforms. ProS$^2$Vi (Qasim and Alisaraie, 2024) not only enhances security but also offers a modern, user-friendly interface and detailed output that leverages the greatest in graphical technology, including 3D icons and high-quality output production for publications or presentations with accessible applications for non-technical biochemists, biologists, or for teaching protein science courses to students.

By running locally, ProS$^2$Vi (Qasim and Alisaraie, 2024) ensures that all the protein data remains under the user's direct control without the need to transmit sensitive information over the internet. This setup is particularly beneficial for handling unpublished or proprietary research data concerning novel protein structures being under investigation *in vitro* using, for instance, protein crystallography and NMR techniques, *in silico* folded protein sequences, or novel bioengineered proteins, all of which require a high level of confidentiality prior to their publication. Furthermore, ProS$^2$Vi (Qasim and Alisaraie, 2024) is designed to be customizable, allowing researchers to adapt and expand its functionalities to meet specific needs. It supports various output file formats, such as Portable Network Graphics (PNG) (Roelofs and Koman, 1999), Joint Photographic Expert Group (JPEG) (Hudson, et al., 2017), and even Portable Document Format (PDF) (Grech, 2002).

## 2. Tool Description

ProS²Vi (Qasim and Alisaraie, 2024) is a 2D secondary assignment depicter tool developed using Python and works using the DSSP algorithm (Kabsch and Sander, 1983) output. It depicts the secondary structure of a protein based on the standard protein folding representative, including α-helices, β-strands, β-bridges, $3_{10}$-helices, H-bonded turns, etc., according to the pertinent descriptions of the DSSP algorithm (Kabsch and Sander, 1983).

*2.1. Implementation*

ProS²Vi (Qasim and Alisaraie, 2024) accepts protein structure with a Protein Data Bank (PDB) (Burley, et al., 2017) "pdb" file format or a macromolecular Crystallographic Information Framework (mmCIF) (Westbrook, et al., 2022) file. It uses Biopython (Cock, et al., 2009) and the DSSP executable file to extract the DSSP output pertaining to the structure of the protein of interest. The DSSP executable file can easily be downloaded on Linux or Windows Subsystem for Linux (WSL) using the standard Advance Packaging Tool (APT) (Project, 1998), which is the default user interface working with core libraries to handle the installation and removal of software on Linux distributions. Next, the DSSP output is formatted in the form of a Python dictionary, where each element represents a separate protein chain — within each protein chain element is a Python list containing amino acid residues and their relevant secondary structure predictions, according to the DSSP output. Once this Python list is ready and includes all the required information about the protein chain identification, amino acid residues, and their relevant secondary structures, it is used to populate a HyperText Markup Language (HTML) table using "Jinja2" (Ronacher, 2008). "Jinja2" is one of the fastest templating engines, which can be used to create HTML documents using Python. Since protein structures can often be quite large, a fast-templating engine is necessary. The result is an HTML table, divided into separate parts by the chains of a multi-subunit protein structure. **(Figure 1)**

The HTML table contains the protein structure file name or its "pdb" ID, from Protein Data Bank, a short description of the protein type, and the organism. The information in the table is followed by the protein's chain ID and its sequence ID from UniProt (i.e., the resource for the protein sequence) (Consortium, 2022), as well as other additional protein functional information.

Each part of the table consists of three alternating rows: *i.,* the first row contains secondary structure indices counting the occurrences of α-helices, β-strands, β-bridges, $3_{10}$-helices, π-helices, H-bonded turns, and simple bends, such as "H1" for the first occurrence of α-helix, "H2" for the second occurrence of α-helix, etc. *ii.* The second row contains the secondary structural icons for residues, exhibiting a coil for α-helices, $3_{10}$-helices, and π-helices with different color representations, an arrow for β-strands and β-bridges, and cylinders of varying thickness for H-bonded turns, simple bends, and unsolved structures and *iii.* The third row consists of single-letter coded amino acids of the protein sequence. **(Figure 2)**

Once the HTML table is created, it is converted into a PDF or rendered as an image. This is carried out via an open-source tool, "wkhtmltopdf" (Truelsen and Kulkarni, 2009), to render HTML documents to PDF and various other image formats. "Wkhtmltopdf" is an open-source command-line tool written in C++ language. Python libraries (e.g., "PDFKit" (JazzCore, 2013) and "IMGKit" (Jarrekk, 2017)) can be utilized to function as wrappers, allowing the use of such tools. **(Figure 1)**

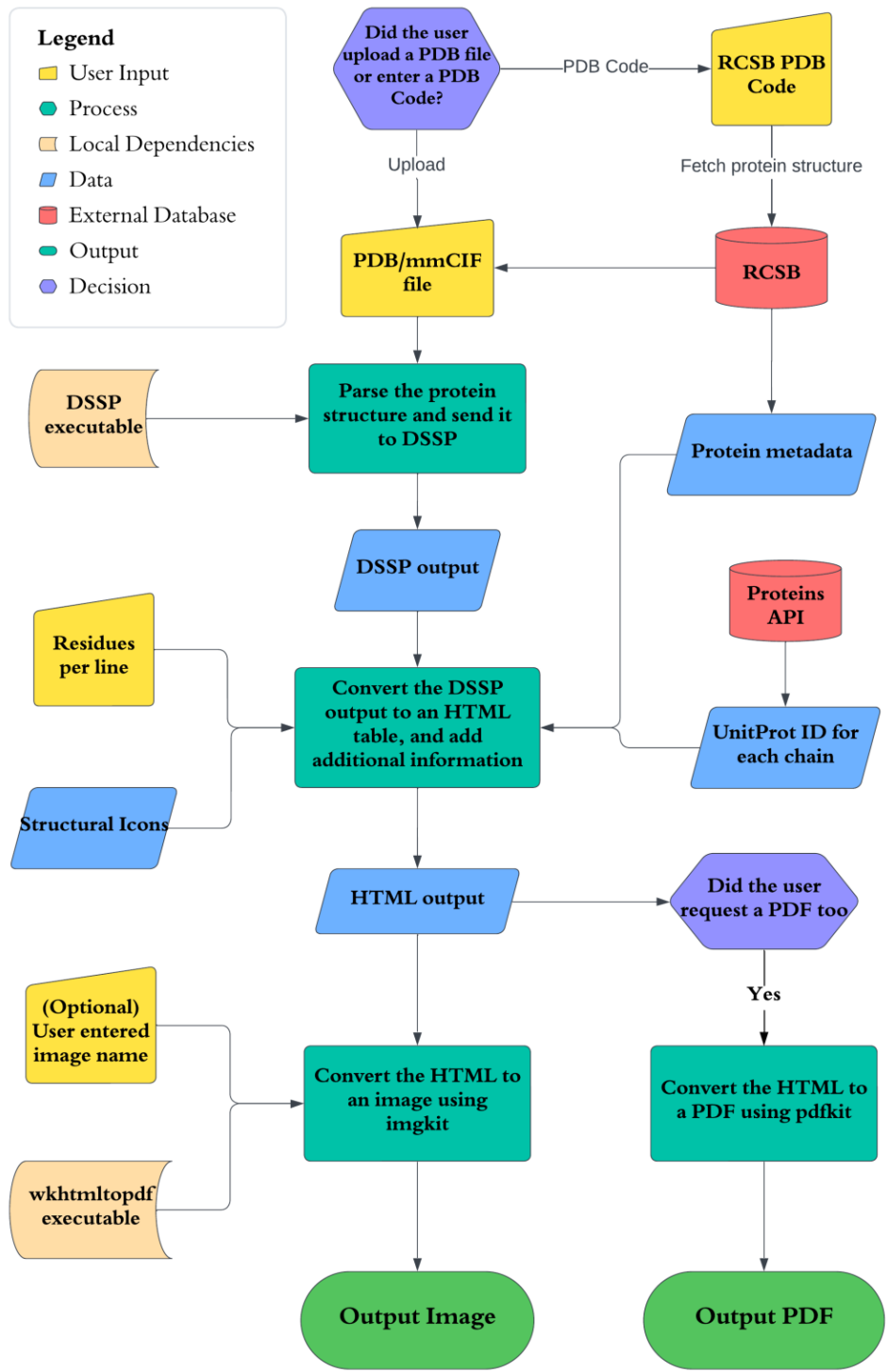

**Figure 1:** The flowchart summarizes ProS$^2$Vi's (Qasim and Alisaraie, 2024) roadmap from the input structure file entry to the generation of the output image.

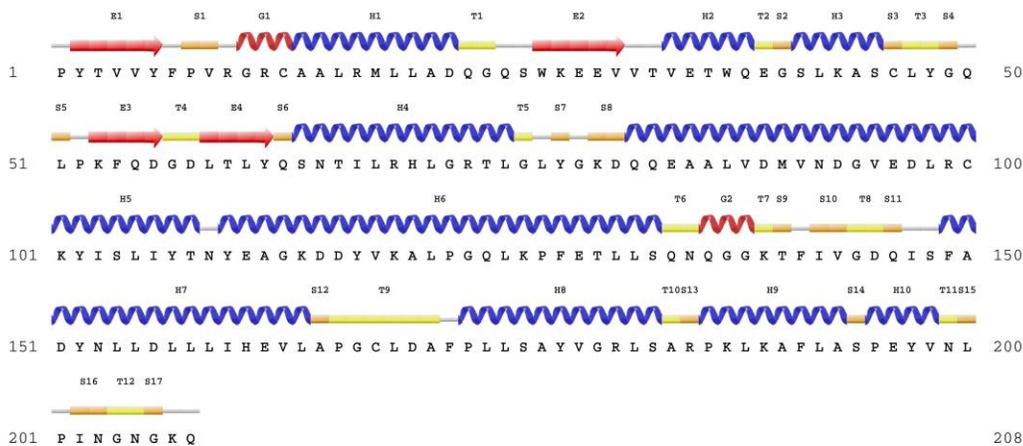

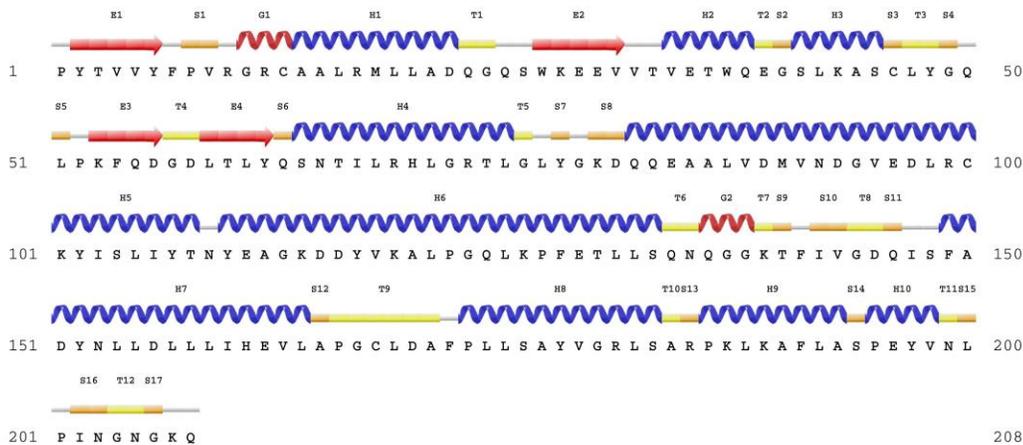

**Figure 2:** An example presenting the visualization for protein structure 1OGS (Oakley, et al., 1997), produced by ProS²Vi (Qasim and Alisaraie, 2024) using default settings. The output visualization clearly shows the file name or protein's PDB code, along with a short informative description of the protein and its sourced organism. All the chains in the multi-subunit protein structure appear along with their amino acid sequences pertaining to the protein identification code (ID) in the database of protein sequences (i.e., UniProt), and relevant secondary structure icons on the top row, as predicted by the DSSP algorithm. This example is generated with default settings and icon colors. *Note that additional options are readily provided to change the number of residues in each row or the color of the icons using the color palette in GUI (See Figure 4).*

*2.2. Requirements*

To use ProS$^2$Vi (Qasim and Alisaraie, 2024), a Linux machine is preferred since its dependencies, including DSSP and "wkhtmltopdf," are much more straightforward to install on Linux than on Windows. To run this on a Windows machine, a Windows Subsystem for Linux (WSL) (Microsoft, 2016) is preferred. WSL allows the users to run a Linux environment on Windows without the need for a separate virtual machine or dual booting. WSL can be downloaded free of charge from Microsoft's official website and is easy to install. Once the environment is set up, the repository containing the code needs to be cloned, and the dependencies need to be installed. The required dependencies to run ProS$^2$Vi (Qasim and Alisaraie, 2024) include Python, DSSP, "wkhtmltopdf," and all other required Python libraries, including "Biopython," "Jinja2", "IMGKit," "PDFKit," "Flask" (Ronacher, 2010), "pdf2image" (Belval, 2017), "CairoSVG" (Kozea, 2011), "Pillow" (Clark, 2010), and "Requests" (Prewitt, et al., 2011). All the Python dependencies are listed in the "requirements.txt" file in the git repository and can be easily installed using the Python package installer (pip). While these dependencies, along with non-Python dependencies like DSSP and "wkhtmltopdf", can be installed manually, we have included a shell script in the package to simplify the installation process for the average user. This shell script examines whether all the required dependencies exist in the system; otherwise, it installs them, making the entire installation process automatic and much easier than manually installing all dependencies individually. Once all the requirements are installed, ProS$^2$Vi (Qasim and Alisaraie, 2024) can be run using either the basic command-line interpreter or the more user-friendly Graphical User Interface (GUI) designed for this SS depicter. **(Figure 3)**

*2.2.1 Difference between Linux and Windows Installation*

The installation of ProS²Vi (Qasim and Alisaraie, 2024) on Linux versus Windows is primarily identical, except for the fact that for Windows, the users first need to download and install WSL, following the steps on Microsoft's official website (Microsoft, 2016). Once that is installed, the user can open the WSL terminal, and the next steps are the same as those for Linux. The next step is cloning the git repository containing the software and installing all the required dependencies, either manually or automatically, by running the commands "chmod +x install_dependencies.sh" and "./install_dependencies.sh". Next, the user can use ProS²Vi's (Qasim and Alisaraie, 2024) GUI by running the command "python3 pros2vi_gui.py" on the terminal, which should open a web browser in Linux. In WSL, users might need to manually open their web browser and enter "http://127.0.0.1:3000" to access the GUI.

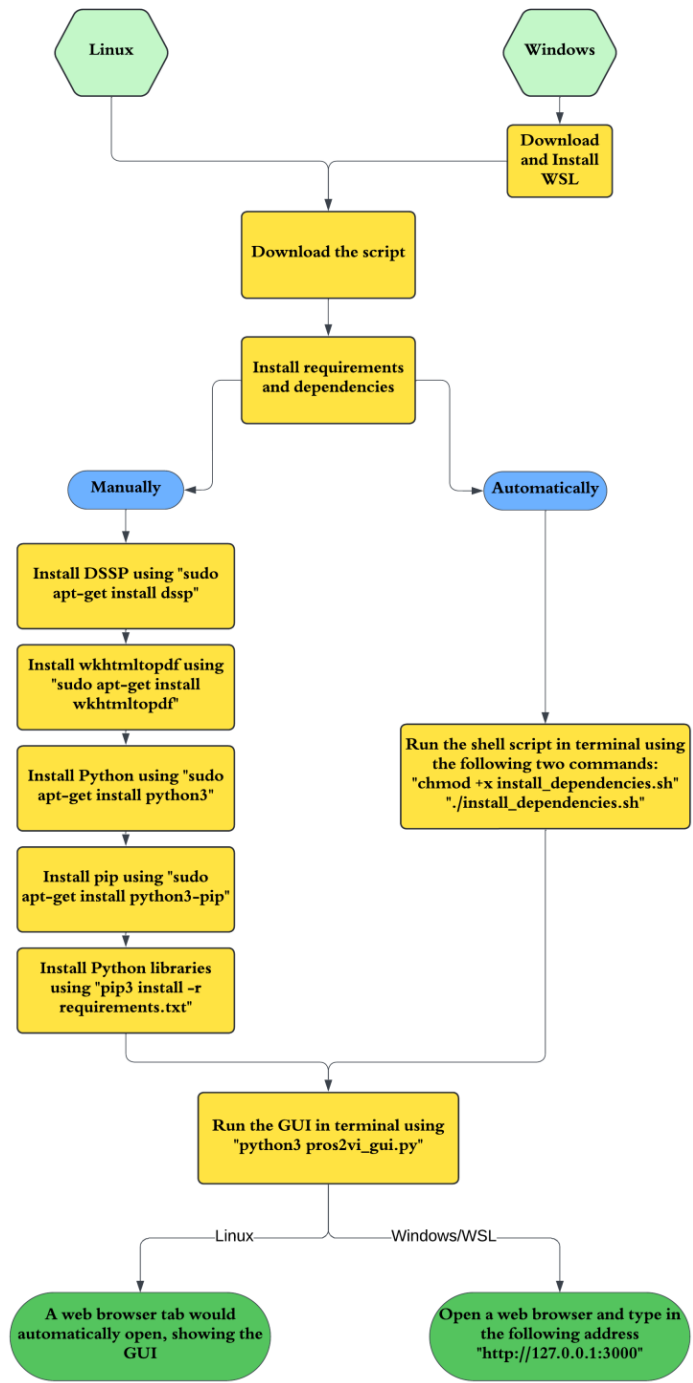

**Figure 3:** The flowchart summarizes the entire installation process for ProS²Vi (Qasim and Alisaraie, 2024) on both Linux and Windows machines. *The use of shell script makes the entire installation process much easier compared to manually installing dependencies.*

*2.3. Graphical User Interface and Customizability*

To use the GUI, a Python file named "pros2vi_gui.py" should be run using the terminal, which then launches an instance of a web browser showing the interface. If the script is run on WSL, the browser might not launch automatically and instead needs to be manually opened by opening a browser and entering the address shown in the terminal subsequent to running the script. It is noteworthy that despite the GUI running on a web browser, it is by no means connected to the internet. **(Figure 4)**

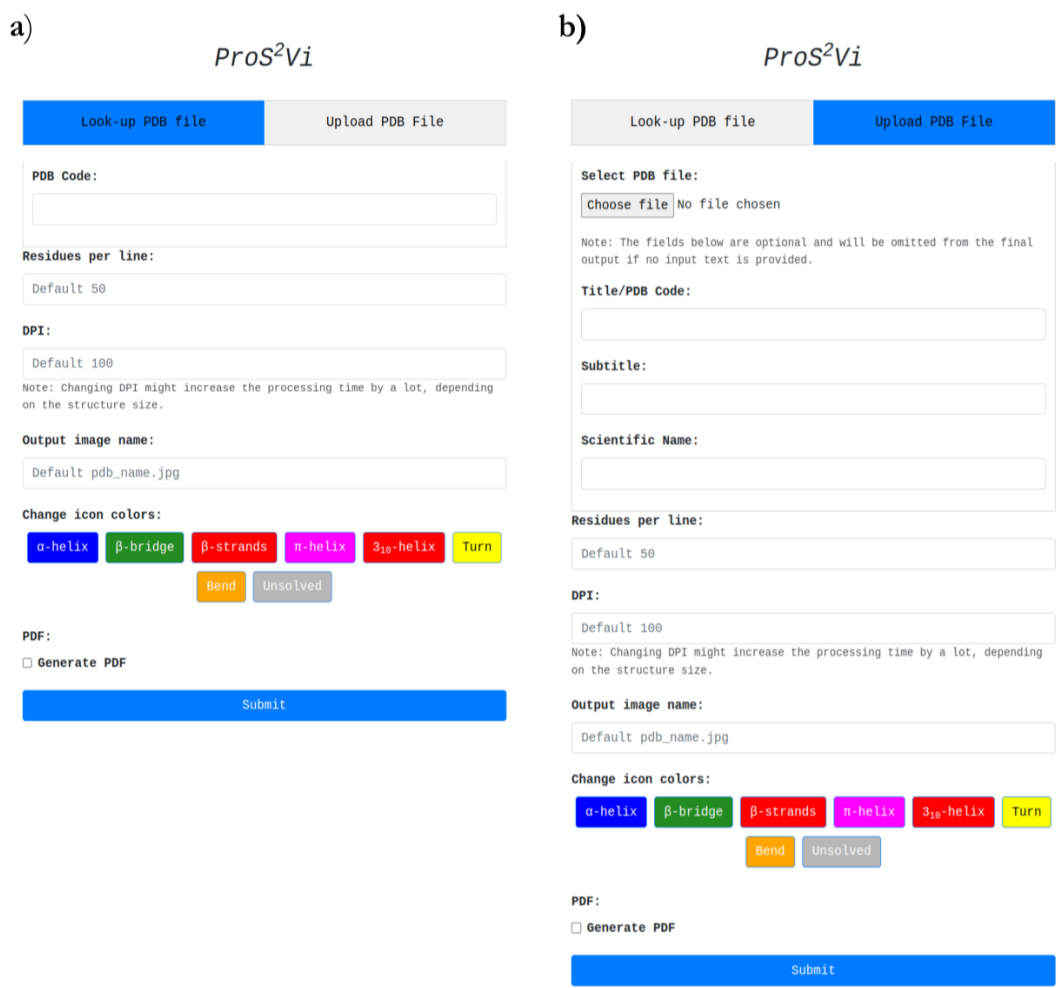

**Figure 4:** The Graphical User Interface (GUI) of ProS²Vi (Qasim and Alisaraie, 2024), demonstrating its layout and the available options. The blue tab at the top shows the current input selection. **A)** the "Look-up PDB file" option can be seen, which only requires an RCSB PDB code

to generate visualization. **B)** the "Upload PDB File" option, which allows the user to upload their own PDB file and add their own "Title," "Subtitle," and "Source Scientific Name". *If either of the "Title," "Subtitle," or "Source Scientific Name" is left empty, then that heading is omitted from the visualization.*

ProS$^2$Vi (Qasim and Alisaraie, 2024) has two input options: "Look-up PDB File" and "Upload PDB File." The first option accepts the protein structure code from the protein data bank (PDB) and downloads the file from the database of the "Research Collaboratory for Structural Bioinformatics" (RCSB) (Burley, et al., 2022). The second option requires the user to upload a file in either "pdb" or the "mmCIF" file format containing the atomic coordination data of any protein's structure. This option is useful for generating visualization for novel protein structures that are not yet available on RCSB. With this option, the users have to manually enter the "Title/PDB Code," "Subtitle," and "Source Scientific Name" of the protein structure. If any of these input boxes are left empty, then the related heading is omitted from the output, providing users with the ability to generate visualizations with no headings.

Once the required structural data inputs are selected, the user can customize the ProS$^2$Vi (Qasim and Alisaraie, 2024) output by entering information in the optional fields. The customizations include the ability to change the number of amino acid residues that appear per line, which defaults to 50. The user can also change the file name of the output image and its format, where the supported formats include JPEG and PNG. If no input is provided in this field, the image name defaults to the protein structure code in a PNG output file format. Furthermore, the ability to easily change the color of structural icons provides flexibility and customization to suit individual needs and preferences. These colors can easily be changed for individual icons by pressing the relevant icon button in the GUI, which then opens a color palette, providing the ability to choose any of the available colors. **(Figure 4)**

The command line interface (CLI) version of this script is less flexible since it does not support the ability to change the colors of the structural icons or access the protein structure files from RCSB; however, it can be useful for technical users because it can be coupled with a shell script to generate batch visualizations. For the CLI, the default settings can easily be modified by providing optional arguments when running the script. More information about the CLI positional arguments can be seen by running the CLI with a "–h" parameter, which represents "–help" and lists all available commands.

The default settings currently produce a PNG image with a 100 dots per inch (DPI) resolution. However, the resolution (DPI) can easily be changed in both GUI and CLI versions. It is noteworthy that increasing the DPI might significantly increase the processing time for generating the visualization or might not work for significantly larger protein structures containing several thousand modeled residues. However, to create a higher resolution for larger structures, a PDF version of the visualization can be produced.

## 3. Assessment

The development of a Python-based visualization tool for protein secondary structures yields several vital outcomes that significantly enhance the analyses and presentation of protein data. We evaluated the performance of ProS$^2$Vi's (Qasim and Alisaraie, 2024) on different protein structures with varying residue counts and chain numbers using "pdb" and "mmCIF" file formats. The following sections provide a detailed analysis of the ProS$^2$Vi (Qasim and Alisaraie, 2024) performance metrics and visualization capabilities.

### 3.1. Extraction and Visualization Accuracy

Assessing ProS$^2$Vi (Qasim and Alisaraie, 2024) showed that it effectively extracted the required information from the input structural files, such as amino acid sequences and their predicted DSSP-based secondary structure. It demonstrated robust performance across various protein structures, accurately presenting secondary structures in a 2D format (e.g., α-helices, β-strands, and other motifs) using distinct icons and colors. This ability to differentiate between various secondary structures is crucial for understanding protein folding and function. The results showcase the capability of ProS$^2$Vi (Qasim and Alisaraie, 2024) to generate precise, high-quality, and detailed visualizations that are easy to interpret. For instance, it allows users to quickly identify structural elements and their spatial relationships within the protein using different colors and icons for various secondary structures. This feature is particularly beneficial for educational purposes and presentations, where self-explanatory, simple, and intuitive visualizations are essential.

*3.2. Performance Metrics*

The efficiency of ProS²Vi (Qasim and Alisaraie, 2024) was evaluated using various protein structures with varying sizes and complexities (e.g., multi-subunit, multi-chain). The performance metrics were recorded for both "pdb" and "mmCIF" file formats and included parsing and extracting DSSP output, creating HTML tables, and converting HTML to images. **(Table 1)**

The time required to parse and extract DSSP output varied significantly between the file formats. For instance, the DSSP parsing time for the "pdb" file of a protein with the retrieving code of 8RJK (Sendker, et al., 2024) was ~58.619 seconds, whereas its "mmCIF" file took ~53.970 seconds. This indicates that "mmCIF" files generally allow faster parsing. The time taken to create HTML tables and convert them to images also varied, however, to a lesser extent. For example, creating an HTML table of a protein with the PDB retrieving code of 2VDC (Cottevieille, et al., 2008) took approximately 32.943 seconds for the "pdb" file format and 31.749 seconds for its "mmCIF."

The conversion of HTML to images was relatively rapid, taking nearly 1.679 to 2.295 seconds across different files. The total time to process and generate visualizations ranged from a few seconds for a short peptide such as 1Q7O (Rienstra, et al., 2002) to over a minute for a multi-chained system such as 8RJK (Sendker, et al., 2024). The faster processing times for "mmCIF" file format highlights the advantage of using this format for large datasets. **(Table 1)**

**Table 1:** Average run time (in seconds) for various protein structures, using both "pdb" and "mmCIF" protein structure file formats.

| PDB Code & sequence length (AA) | | Parsing and extracting DSSP output (s) | Creating HTML table (s) | Converting HTML to image (s) | Total time (s) |
|---|---|---|---|---|---|
| **1Q7O** 3 | pdb | 0.441 | 0.035 | 0.328 | 0.805 |
| | mmCIF | 0.241 | 0.034 | 0.323 | 0.597 |
| **4DX9** 3,767 | pdb | 6.437 | 4.283 | 0.545 | 11.264 |
| | mmCIF | 4.310 | 4.223 | 0.535 | 9.068 |
| **3KGV** 4,064 | pdb | 10.601 | 11.098 | 0.790 | 22.489 |
| | mmCIF | 9.447 | 10.608 | 0.748 | 20.804 |
| **2VDC** 11,568 | pdb | 30.481 | 32.943 | 2.295 | 65.721 |
| | mmCIF | 23.894 | 31.749 | 1.679 | 57.323 |
| **8RJK** 19,548 | pdb | 58.619 | 28.208 | 1.510 | 88.346 |
| | mmCIF | 53.970 | 27.914 | 1.487 | 83.371 |
| **8VRJ** 24,650 | pdb | - | - | - | - |
| | mmCIF | 91.443 | 29.923 | 1.658 | 123.025 |

*3.3. Case Study Examples*

As mentioned, the visualization for 8RJK, a large protein with 19,548 modeled residues, demonstrated ProS²Vi's (Qasim and Alisaraie, 2024) robustness in processing complex datasets. The total processing time for the PDB file was ~88.346 seconds, with most of the time (~58.619 seconds) spent on DSSP parsing. The "mmCIF" file processed faster at ~83.371 seconds, showcasing the efficiency gains with this format. Furthermore, the visualization for 8VRJ (Aher, et al., 2024), the largest size protein structure available to test, with 24,650 modeled residues, required ~123.025 seconds to process, with most of the time (~91.443 seconds) taken by DSSP calculation. This test further proves the efficiency of our code, considering the sheer size of the protein structure. For a short peptide such as 1Q7O, with only a few characterized residues, the tool processed the PDB file in ~0.805 seconds and the mmCIF file in ~0.597 seconds. This case highlights the ability of ProS²Vi (Qasim and Alisaraie, 2024) to treat minor-size datasets quickly; however, it is also well-functional for a wide range of protein sizes. The results demonstrate that the Python-based visualization tool is highly effective in extracting, processing, and visualizing protein secondary structures. Its performance across different protein sizes and formats showcases its versatility and robustness.

*3.4. Security and Accessibility*

ProS²Vi (Qasim and Alisaraie, 2024) operating locally eliminates the risks associated with uploading sensitive data to online servers, ensuring the confidentiality and integrity of proprietary research data. That is essential for handling unpublished novel protein structures. Local operation means that sensitive data remains on the user's machine, reducing the risk of exposure to cyber threats that are prevalent on web-based platforms.

*3.5. Performance and File Compatibility*

The tool supports multiple file formats, including "pdb" and "mmCIF" ensuring broad compatibility with existing bioinformatics workflows and databases. This flexibility allows users to work with various data sources without the need for extensive data conversion. The ability to process both file formats is crucial because it will enable researchers to choose the format that best suits their needs, considering that "mmCIF" files often provide slightly faster parsing times than their equivalent "pdb" counterparts. **(Table 1)**

The length of the protein sequences can significantly impact performance, potentially increasing processing times for extensive datasets. This is an essential consideration for researchers working with extensive data, affecting the tool's efficiency and the ability to deliver fast output. ProS²Vi (Qasim and Alisaraie, 2024) was initially created using "WeasyPrint" (Sapin, et al., 2011) and "pdf2image", instead of "wkhtmltopdf", "PDFKit" and "IMGKit" to eliminate the dependency on "wkhtmltopdf", and to allow it to generate images with higher DPI. This implementation was later changed to use "wkhtmltopdf" instead because "WeasyPrint" and "pdf2image" took a considerable amount of time to convert the HTML table to a PDF or image, which increased significantly with the proteins with a greater number of amino acids or more significant structure size. That was tested by comparing the two implementations and the time required to visualize specific protein structures using the "mmCIF" file format. The time necessary for "WeasyPrint" is significantly higher than that for "wkhtmltopdf." For 3KGV (Sibanda, et al., 2010), which contains 4,064 modeled residues, "WeasyPrint" takes around 245.437 seconds to convert the HTML to an image, whereas "wkhtmltopdf" takes only ~0.748 seconds, which shows a significant improvement. **(Table 2)**

**Table 2:** The comparison between the necessary time for "WeasyPrint" and "wkhtmltopdf" (in seconds) to convert an HTML table to an image.

| PDB Code | Time to convert HTML to Image (s) | |
|---|---|---|
| | Using "WeasyPrint" and "pdf2iamge" | Using "wkhtmltopdf" and "imgkit" |
| **1Q7O** | 0.458 | 0.323 |
| **4DX9** | 70.892 | 0.535 |
| **3KGV** | 245.437 | 0.748 |

*3.6. Limitations*

While ProS$^2$Vi (Qasim and Alisaraie, 2024) offers significant advantages, there are limitations to consider. Currently, the tool can effectively process protein structures with up to ~25,000 modeled residues due to the restrictions imposed by the ability of the DSSP algorithm and Biopython. This constraint means that extremely large-sized proteins may require alternative methods for secondary structural assignments. Additionally, the faster processing times for ProS$^2$Vi (Qasim and Alisaraie, 2024) currently only work when producing images with a 100 DPI resolution due to the limitations of "wkhtmltopdf". When creating an image with a different DPI, the algorithm first produces a PDF and then converts it into an image using "pdf2image" instead of "IMGKit," which slightly increases the processing times for smaller protein structures. However, for much larger protein structures, the time increases significantly, and some significantly large structures (>~6,000 modeled residues) might not produce any output at all, depending on the machine specifications. Addressing these limitations in future versions of the tool will be a priority to enhance its applicability and performance further.

*3.7. Customizability and Comparison*

The tool features a contemporary design with customizable icon colors and the ability to export visualizations as scalable PDFs. Modern elements of the GUI ensure that users can intuitively navigate the application, customize their visualizations according to their preferences, and produce high-quality outputs suitable for presentations and publications. Furthermore, the ability to directly download a protein structure from RCSB using the PDB code adds to the accessibility and ease of use. Even though ProS²Vi (Qasim and Alisaraie, 2024) can connect to RCSB to download protein structures, any uploaded protein structure files are still stored and used locally, keeping security a top priority.

This modern interface not only improves the aesthetic appeal but also enhances functionality, making the tool more approachable and easier to use compared to existing solutions such as POLYVIEW-2D (Porollo, et al., 2004) and STRIDE (Heinig and Frishman, 2004). Neither of these tools accepts an mmCIF file format, which is essential because some of the larger protein structures are only supported by the mmCIF file format. Furthermore, neither of the tools provides the option to change or modify the color of structural icons, and additionally, STRIDE is unable to offer the option to save or download a high-quality image of the visualization easily. Most importantly, none can be used locally to generate a protein's secondary structure diagrams. Our tool also automatically saves the generated image and can also create a PDF document. Additionally, ProS²Vi's (Qasim and Alisaraie, 2024) visualization also includes the protein structure's title and source scientific name with the help of the RCSB application programming interface (API) (Rose, et al., 2021), UniProt ID for each chain using the Proteins API (Nightingale, et al., 2017), and a legend (i.e., key).

For the technical users, the output can be further customized by adding changes to the accompanying Cascading Style Sheets (CSS) according to user preference. This file is named "output_styles.css" and is located inside the "templates" folder. To change the font style, the value of the "font-family" property inside the "body" selector can be changed to any other font supported by the web browser. Font size can also be easily changed by modifying the values of the "font-size" appropriately for different parts of the output such as "title," "legend," etc.

## 4. Conclusion

ProS²Vi (Qasim and Alisaraie, 2024), a Python-based visualization tool, offers a secure, advanced, and user-friendly solution for visualizing protein secondary structures. Its local operation provides enhanced security and control, while the modern interface and broad file format compatibility make it a versatile tool for researchers and educators alike. An intuitive graphical user interface allows users to easily navigate the tool and perform complex analyses without requiring extensive coding skills or computational background, enabling a broader range of users, from students to researchers, to benefit from this tool. The GUI is designed to streamline workflow, allowing users to load protein structures, process them through the DSSP algorithm, and visualize the results with minimal steps. ProS²Vi's (Qasim and Alisaraie, 2024) ability to export visualizations as scalable PDFs is a significant advantage, allowing users to produce high-quality images. This feature ensures that the visualizations can be scaled up for presentations and publications without losing image quality. The ability to produce precise, customizable visualizations and export them as scalable PDFs adds significant value, making ProS²Vi (Qasim and Alisaraie, 2024) a valuable resource for researchers and educators. ProS2Vi's (Qasim and Alisaraie, 2024) performance assessment highlights its efficiency. Comprehensive testing and user feedback will guide future enhancements, ensuring the tool remains an asset in the field of structural biology and bioinformatics.

The software is readily available at https://github.com/Alisaraie-Group/ProS2Vi (DOI: https://zenodo.org/records/12554831).

**Associated Content**

*Data Availability Statement*

The ProS²Vi (Qasim and Alisaraie, 2024) source code is freely available at https://github.com/Alisaraie-Group/ProS2Vi (DOI:10.5281/zenodo.12554830) under an open-source Apache 2.0 license.

**Author Information**

*Corresponding Authors*

Laleh Alisaraie − School of Pharmacy, Memorial University of Newfoundland, 300 Prince Philip Dr, A1B 3V6, St. John's, Canada; Email: laleh.alisaraie@mun.ca

*Author Contributions*

The manuscript was written with the contributions of all authors, and all authors have approved the final version.

**Conflict of Interest**: The authors declare no conflict or competing financial interest.

**Funding Sources**

This work was supported by funding from the Natural Sciences and Engineering Research Council of Canada (NSERC), Discovery Grant (No. 212654), awarded to LA.

Grech, V. The Portable Document Format - PDF. *Images in paediatric cardiology* 2002;4(2):1-3.

Gu, Q. and Liu, P. Denial of Service Attacks. In, *Handbook of Computer Networks*. John Wiley and Sons; 2012. p. 454-468.

Heinig, M. and Frishman, D. STRIDE: a web server for secondary structure assignment from known atomic coordinates of proteins. *Nucleic acids research* 2004;32(Web Server issue):W500-502.

Hudson, G.*, et al.* JPEG at 25: Still Going Strong. *IEEE MultiMedia* 2017;24(2):96-103.

Jarrekk. 2017. IMGKit: Python library of HTML to IMG wrapper. https://github.com/jarrekk/imgkit

JazzCore. 2013. Python-PDFKit: HTML to PDF wrapper. https://github.com/JazzCore/python-pdfkit

Kabsch, W. and Sander, C. Dictionary of protein secondary structure: pattern recognition of hydrogen-bonded and geometrical features. *Biopolymers* 1983;22(12):2577-2637.

Koch, O. Use of secondary structure element information in drug design: polypharmacology and conserved motifs in protein-ligand binding and protein-protein interfaces. *Future medicinal chemistry* 2011;3(6):699-708.

Kozea. 2011. CairoSVG. https://cairosvg.org/

Martin, J.*, et al.* Protein secondary structure assignment revisited: a detailed analysis of different assignment methods. *BMC Structural Biology* 2005;5(1):17.

Microsoft. 2016. Windows Subsystem for Linux. https://learn.microsoft.com/en-us/windows/wsl/install

Nawaz, N.A.*, et al.* A comprehensive review of security threats and solutions for the online social networks industry. *PeerJ. Computer science* 2023;9:e1143.

Nightingale, A.*, et al.* The Proteins API: accessing key integrated protein and genome information. *Nucleic acids research* 2017;45(W1):W539-W544.

O'Donoghue, S.I. Grand Challenges in Bioinformatics Data Visualization. *Frontiers in bioinformatics* 2021;1:669186.

Oakley, A.J.*, et al.* The structures of human glutathione transferase P1-1 in complex with glutathione and various inhibitors at high resolution. *J Mol Biol* 1997;274(1):84-100.

Porollo, A.A., Adamczak, R. and Meller, J. POLYVIEW: a flexible visualization tool for structural and functional annotations of proteins. *Bioinformatics (Oxford, England)* 2004;20(15):2460-2462.